\documentclass[prl,twocolumn,aps,showpacs,preprintnumbers]{revtex4}
\usepackage{graphicx}
\usepackage{dcolumn}
\usepackage{bm}

\begin{document}

\title{Erratum: Single-electron Tunneling with Strong Mechanical Feedback}

\author{Ya.~M.~Blanter, O.~Usmani, and Yu.~V.~Nazarov} 
 \affiliation{Kavli Institute of Nanoscience, Delft University of
 Technology, Lorentzweg 1, 2628 CJ Delft, The Netherlands}

\date{\today}

\maketitle

We have discovered an unfortunate error in our recently published article 
\cite{previous04}. Namely, we erroneously omit a term
in the right-hand side of Eq. (5), that reads
\begin{equation}
\frac{F}{M \Gamma_t^3}\left( \Gamma^{0\to 1} \frac{\partial}{\partial
x} \Gamma^{1\to 0} - \Gamma^{1\to 0} \frac{\partial}{\partial 
x} \Gamma^{0\to 1} \right) \frac{\partial}{\partial v} (vP).  
\end{equation} 

This term has a clear physical meaning
of a  friction force 
acting on mechanical part of the system. This force arises from
electron tunneling and efficiently dumps the oscillator. 
For a tunneling through a single level, this force is
proportional to $\delta(\tilde W - Fx)$ and only occurs once the
amplitude is so big that the oscillator is driven below the current
threshold, this was a reason for erroneous omitting of this term.
However, the friction is effective at the threshold. 
As a result, the distribution function $P(E)$ for
infinitely large quality factor $Q$, is constant for $E < \tilde W^2
M\omega_0^2/F^2$ and decreases rapidly for higher energies. In other
words, the oscillation amplitude $\zeta$ grows as far as $\hat W -
F\zeta$ does not reach the threshold, and is stabilized
afterwards. The scale $W_c$ is determined by this dissipation term,
and happens to be $F^2/\omega_0^2 M = \hbar \lambda \omega_0 \ll \hbar
\omega_0$. For such small scales our classical description is not
applicable, and thus the mechanical motion in this regime does not
renormalize transport properties of a SET device. Similar conclusions
are drawn for tunneling through continuum of levels. We conclude
that the regime of strong mechanical feedback described in the article
never takes place for the models considered.

We are confident, however, that this regime certainly occurs 
for more realistic models that take into account energy dependence
the tunneling matrix elements. 
Strong current renormalization and
enhanced current noise typical for this regime can be described within the same
formalism as in the article provided the
friction force is taken  into account. 
Quantitative results  will be presented elsewhere.

\end{document}